%%%%%%%%%%%%%%%%%%%%%%  filename = paper.lanl.ellis.tex  %%%%%%%%%%%%%%%%%%%%%%
%%%%%%                                                                   %%%%%%
%%%%%%  This source file is written in AMSTeX.  There are five figures,  %%%%%%
%%%%%%  called in from eps files with the \epsffile command.             %%%%%%
%%%%%%                                                                   %%%%%%
%%%%%%%%%%%%%%%%%%%%%%%%%%%%%%%%%%%%%%%%%%%%%%%%%%%%%%%%%%%%%%%%%%%%%%%%%%%%%%%

\input amstex
\input epsf

\def\[[{{[\hskip -1.5pt [}}         % the notation for dbl. left ['s
\def\]]{{]\hskip -1.5pt ]}}         % the notation for dbl. rght ]'s

\def\degree{\mathaccent"17 {}}

\documentstyle{amsppt}
\loadbold
\loadeusm
\TagsOnRight
\NoBlackBoxes

\nologo

\magnification=\magstephalf

\hoffset=.25truein
\hsize=6.0truein
\vsize=9.0truein
\parindent=20pt

\topmatter
\title An Expanding Universe of Spinning Spheres \endtitle
\author Homer G. Ellis\endauthor

\abstract{\it {A novel but elementary geometric construction produces on the
seven-dimensional manifold of rotated spheres in Euclidean three-space a
finslerian geometry whose geodesics are interpreted as the paths of free,
spinning, spherical particles moving through de Sitter's expanding universe.
A particle of nonzero inertial rest mass typically follows a helical track and
exhibits behavior remindful of the phenomenon of ``Zitterbewegung'' of spinning
electrons first deduced by Schr\"odinger from Dirac's relativistic wave
equation.  Its velocity vector and its spin vector precess about the axial
direction of the helix, with their projections onto that direction at all times
parallel or at all times antiparallel.  Particles of zero rest mass follow
straight tracks at the speed of light with their spin vectors parallel or
antiparallel to their velocity vectors, thereby replicating behavior of
spinning photons predicted by the quantum theory of light.}}
\endabstract

\endtopmatter

\document

The four-dimensional manifold whose points are the spheres of Euclidean
three-space $\Bbb E^3$ can be coordinatized by $\[[\, R, \bold s \,\]]$, this
designating the sphere $S$ of radius $R$ with its center $C$ at position
$\bold s$.  If $\[[\, R + dR, \bold s + \bold {ds} \,\]]$ designates a
neighboring sphere $S'$, and $d \alpha$ is the radian measure of the angle in
which $S$ and $S'$ intersect, then
$$
\aligned
d\alpha^2 &= (ds^2 - dR^2)/@!@!@!@!R^2  \\
          &= e^{2 t} ds^2 - dt^2,
\endaligned
\tag1
$$
where $ds := |\bold {ds}|$ and $t := -\ln R$ (see Fig\. 1).  As this is
precisely the metric of de Sitter's expanding universe, one can consider that
universe to {\it be} this manifold of spheres, the event at
$\[[\,t, \bold s\,\]]$ in de Sitter's universe being then the two-sphere of
radius $e^{-t}$ centered at position $\bold s$ in $\Bbb E^3$.  One gains
thereby the advantage of reducing the ever mysterious notion of time ($t$) to a
purely spatial concept ($-\ln R$), along with the satisfaction of producing a
space-time cosmological model out of the whole cloth of Euclidean
space.$^{(1)}$  This satisfaction is tempered, however, by the apparent absence
of a way to extend the construction to a metric for spheres that are
``spinning'' in a sense that makes sense.  The difficulty lies in the fact that
neighboring spheres will intersect in the same angle whether spinning or not.

A plan of escape from this cul-de-sac grows out of the realization that
radian measure of an angle is simply a ratio of arc lengths, which suggests
that some alternative characterization of $d \alpha$ as a ratio of distances
might admit the needed extension.  Of several such characterizations, the one
that does the job is this: If each point $P$ of the sphere $S$ is moved
radially, to produce a magnification of $S$ by the factor
$1 + dR@,@,@,/@!@!@!@!R$, and subsequently is translated by the vector
$\bold {ds}$, then $P$ arrives at a point $P'$ on the neighboring sphere $S'$.
Generically, there are only two such points $P$ for which the displacement
vector $\overarrow {PP'}$ is orthogonal to $S$, namely, the two points where
the line through $C$ and the center $C'$ of $S'$ intersects $S$.  Of these two
points $P$ one has moved a distance $dR + ds$ in the direction of
$\overarrow {CP}$, the other a distance $dR - ds$ in the direction of
$\overarrow {CP}$.  The product of the ratios of these distances to $R$ is
exactly the negative of the $d \alpha ^2$ of Eq\. (1), even when, as in
Fig\. 2, $S$ and $S'$ fail to intersect, so that there is no angle to measure.

\vskip 15pt

\epsfxsize=5truein
\epsffile[25 430 537 721]{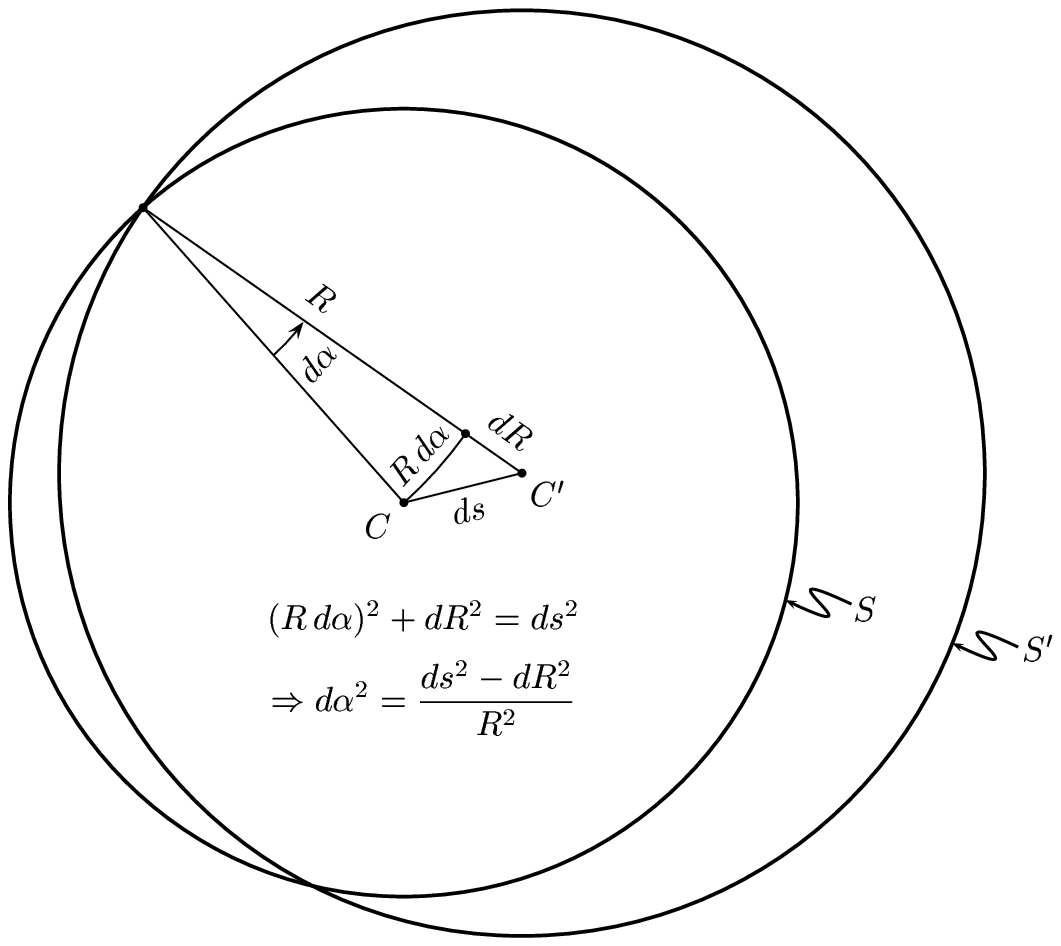}

\midspace{-7pt} \caption{Fig\. 1.   Intersecting neighboring spheres $S$ and
$S'$ in Euclidean three-space, with angular separation $d \alpha$ --- shown in
cross section through their centers.}

Now the means of escape is at hand.  It is to include a rotation with the
magnification and the translation, find the points $P$ of $S$ for which
$\overarrow {PP'}$ is orthogonal to $S$, compute the ratios to $R$ of distances
moved, as before, and use these to define a measure of the separation of $S'$
from $S$.  A hurdle or two remain, however.  The first is the need to specify
the manifold that will take the role played by the sphere manifold in the
nonrotating case.  Clearly, this will be the manifold of rotat{\it ed\/}
spheres, a point of which is a sphere with a center, a radius, and a rotational
position relative to some standard position for all spheres with the same
center.  This seven-dimensional manifold $\eusm M$, diffeomorphic to
$\Bbb R \times \Bbb E^3 \times \text{SO}(3)$, can be coordinatized by
$\[[\, R, \bold s, \phi, \theta, \psi \,\]]$, where $\phi$, $\theta$, and
$\psi$ are Euler angles that together specify the rotational position of the
sphere with respect to the standard reference frame at~$\bold s$.  With
$t := -\ln R$ as before, a path in this manifold can be taken to represent a
spherical particle, moving through space and time, spinning as it goes.

Let the rotated sphere $S$ designated by
$\[[\, R, \bold s, \phi, \theta, \psi \,\]]$ undergo the combined infinitesimal
rotation, expansion, and translation represented by
$\[[\, dR, \bold {ds}, d\phi, d\theta, d\psi \,\]]$.  Let $P$ be a point of
$S$, and let $\bold u = \overarrow {CP}$, the position vector of $P$ relative
to the center $C$ of~$S$.  Then the rotation moves $P$ to a point whose
position vector relative to $C$ is
$\bold u + \boldsymbol\delta \times \bold u$, where
$\boldsymbol\delta := \[[\, (\cos \phi)
d\theta + (\sin \phi)(\sin \theta) d\psi, (\sin \phi) d\theta -
(\cos \phi)(\sin \theta) d\psi, d\phi + (\cos \theta) d\psi \,\]]$.
The magnification multiplies this vector by $1 + dR@,@,@,/@!@!@!@!R$, and the
translation adds $\bold {ds}$.  Thus the requirement that the final position
$P'$ of $P$ be collinear with $C$ and $P$, equivalent to the requirement that
$\overarrow {PP'}$ be orthogonal to $S$, reduces to the equation
$$
(1 + dR@,@,@,/@!@!@!@!R)(\bold u
   + \boldsymbol\delta \times \bold u) + \bold {ds} =
(1 + \rho)\bold u,
\tag2
$$
for some number $\rho$.  When the term of second order in the infinitesimals is
discarded, this equation simplifies to
$$
(\rho - dR@,@,@,/@!@!@!@!R) \bold u =
\boldsymbol\delta \times \bold u + \bold {ds}.
\tag3
$$
In the generic case that $\Delta :=
(\rho - dR@,@,@,/@!@!@!@!R)[(\rho - dR@,@,@,/@!@!@!@!R)^2 + \delta^2] \neq 0$,
the solution of this equation is
$$
\bold u = [(\rho - dR@,@,@,/@!@!@!@!R)^2 \bold {ds}
           + (\rho - dR@,@,@,/@!@!@!@!R)(\boldsymbol\delta \times \bold {ds})
           + (\boldsymbol\delta \cdot \bold {ds})\boldsymbol\delta]/\Delta.
\tag4
$$
Because $P$ lies on $S$, $\bold u \! \cdot \! \bold u = R^2$, which is
equivalent to
$$
(\rho - dR@,@,@,/@!@!@!@!R)^4
 + [\delta^2 - (ds/@!@!@!@!R)^2](\rho
             - dR@,@,@,/@!@!@!@!R)^2
             - [\boldsymbol\delta \cdot (\bold {ds}/@!@!@!@!R)]^2 = 0.
\tag5
$$
The numbers $\rho$ that satisfy this equation are the distance ratios with
which to build a separation measure on $\eusm M$ and make good our escape.
Generically, there are four such numbers, two of them real, the others complex.
The remaining hurdle is to decide how best to use them.

\vskip 15pt

\epsfxsize=5truein
\epsffile[70 410 511 761]{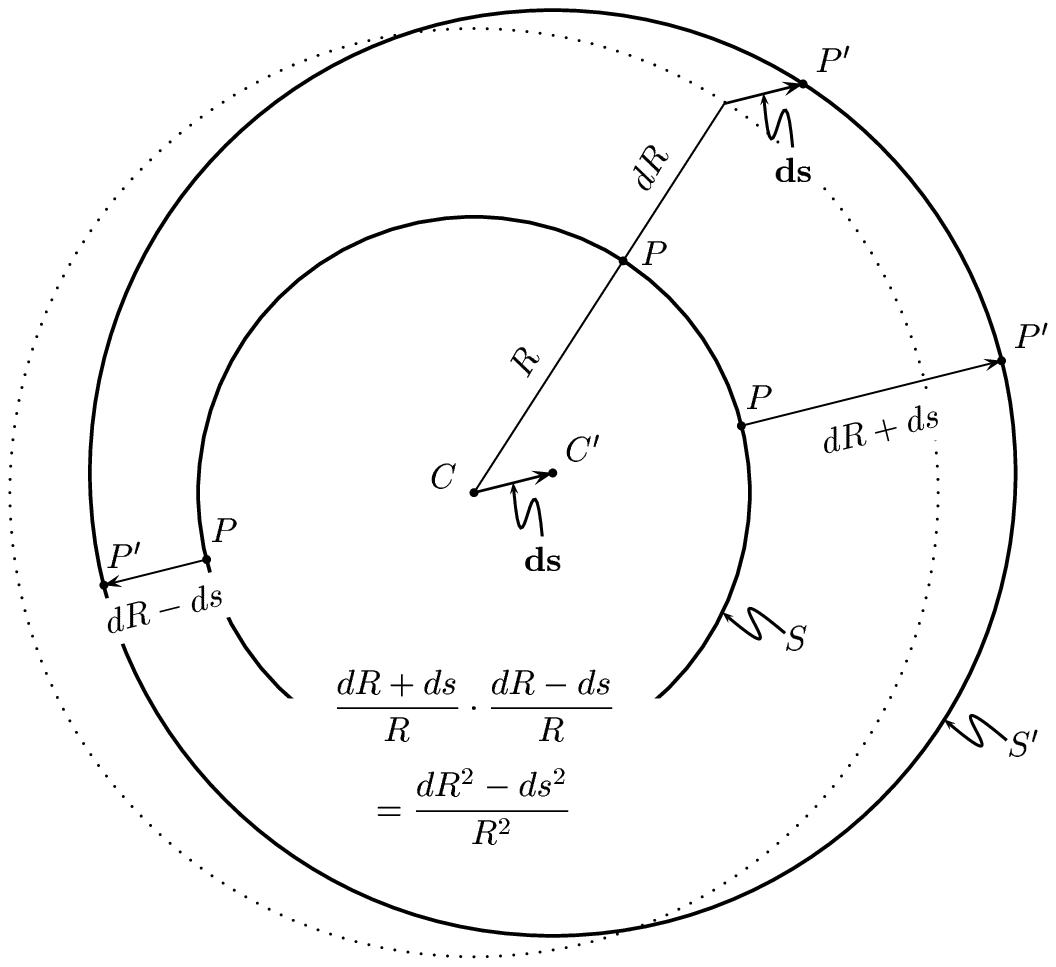}

\midspace{0pt} \caption{Fig\. 2.   Nonintersecting neighboring spheres $S$ and
$S'$ in Euclidean three-space, separated by the ``two-point'' distance
$\sqrt{(dR^2 - ds^2)/@!@!@!@!R^2}$ --- shown in centered cross section.}

Inasmuch as only the two real roots of Eq\. (5) correspond to real points of
$S$, one might think it best to construct the separation measure from their
product.  Investigation shows, however, that this yields a measure with an
incurable degeneracy.  If on the other hand one chooses to build the separation
measure from the product of all four of these ratios, then smooth sailing lies
ahead, but no longer on the broad Sea of Riemann, rather on the vaster Ocean of
finslerian Geometry.

A finslerian geometry on a manifold such as $\eusm M$ assigns to each smooth
path $p\: \botsmash{[}a, b\botsmash{]} \to \eusm M$ an integrated length
$I(p) := \int_a^b L(p, \dot p)$, subject for present purposes essentially only
to the restrictions that $L$ be positively homogeneous of degree one in the
velocity $\dot p$ (so that $I(p)$ will be independent of path parametrization)
and that the metric tensor $G$, loosely described as
$dx^M \otimes g_{MN} dx^N$, where
$g_{MN}(x, v) := \partial^2 [(1/2)L^2(x, v)] / {\partial v^M \partial v^N}$
if $x = \[[\, x^K \,\]]$ and $v = v^K (\partial/ {\partial x^K})$, be
nondegenerate.  The homogeneity allows the finslerian metric function $L$ to be
reconstructed from $G$ via the equation $L^2(x, v) = v^M g_{MN}(x, v) v^N$.
Riemannian geometries are those finslerian geometries for which $g_{MN} (x, v)$
is independent of $v$.$^{(2,3)}$

The product of the four roots of Eq\. (5) is expressible both as
$$
dt^2 (dt^2 - e^{2 t} ds^2 + \delta^2) -
e^{2 t} (\bold {ds} \cdot \boldsymbol \delta)^2
\tag6
$$
\vskip -15pt
\noindent and as
$$
(dt^2 + \delta^2)(dt^2 - e^{2 t} ds^2) +
e^{2 t} | \bold {ds} \times \boldsymbol \delta |^2,
\tag7
$$
where again $ t:= -\ln R$.  With this product we can now impress upon $\eusm M$
a finslerian geometry by defining $L$ as follows:  Let
$\[[\,\, p^K \,\]] := \[[\, x^K(p) \,\]]$,
$= \[[\, t, \bold s, \phi, \theta, \psi \,\]]$ for short, and
$\[[\,\, {\dot p}^K \,\]]
:= \[[\, dx^K(p)\dot p \,\]] = \[[\, (p^K)\,\dot{} \,\,\]]$,
$= \[[\, \dot t, \dot \bold s, \dot \phi, \dot \theta, \dot \psi \,\]]$
for short.  Let $\boldsymbol \sigma := \[[\, (\cos \phi) \dot \theta +
(\sin \phi)(\sin \theta) \dot \psi, \allowmathbreak
(\sin \phi) \dot \theta -
(\cos \phi)(\sin \theta) \dot \psi, \dot \phi + (\cos \theta) \dot \psi \,\]]$.
Then
$$
\aligned
L(p, \dot p)
 :\!@!@!@!@!@!@! &= \big| \dot t^2 (\dot t^2 - e^{2 t} \dot s^2 + \sigma^2) -
                             e^{2 t} (\dot \bold s \cdot
                                      \boldsymbol \sigma)^2 \big|^{1/4}  \\
                 &= \big| (\dot t^2 + \sigma^2)(\dot t^2 - e^{2 t}
                                      \dot s^2) + e^{2 t} | \dot \bold s \times 
                                      \boldsymbol \sigma |^2 \big|^{1/4}.
\endaligned
\tag8
$$
As is seen most clearly in (7) above, this finslerian geometry incorporates the
geometrically derived de Sitter space-time metric of Eqs\. (1) and envelops it
in additonal structure involving the rotations of the spheres.  The geodesic
paths of the finslerian geometry will be taken to represent freely spinning
spherical particles moving through de Sitter's universe under the influence
only of the gravitational effects attributable to the cosmic expansion.

Computing the Euler--Lagrange equations for stationary paths of $I$, and
applying to them the inverse of the metric tensor $G$ to isolate the
derivatives, one arrives at the following equations for affinely parametrized
geodesics:
$$
\align
                   \dot E &= C_0,
\tag9  \\
             \dot \bold v &= C_1 \bold v + C_2 \boldsymbol \sigma +
                             C_3 (\bold v \times \boldsymbol \sigma),
\tag10 \\
\dot {\boldsymbol \sigma} &= C_4 \bold v + C_5 \boldsymbol \sigma +
                             C_6 (\bold v \times \boldsymbol \sigma),
\tag11
\endalign
$$
in which $E := \dot t, \bold v := e^t \dot \bold s$, and
$$
\aligned
C_0 &= -(E^2 v^2 - B^2)/D,  \\
C_1 &= -E[(E^4 + B^2)(E^2 - v^2) + (E^4 - B^2)(E^2 - \sigma^2)]/
                                                      D(E^4 + B^2),  \\
C_2 &= -2B E(E^2 v^2 - B^2)/D(E^4 + B^2),  \\
C_3 &= -B^2/(E^4 + B^2),  \\
C_4 &= -2 BE[(E^4 + B^2) + E^2(E^2 - \sigma^2)]/D(E^4 + B^2),  \\
C_5 &= 2E^3(E^2 v^2 - B^2)/D(E^4 + B^2),  \\
C_6 &= B E^2/(E^4 + B^2),
\endaligned
\tag12  
$$
with $B =
\bold v \! \cdot \! \boldsymbol \sigma$ and $D = 2E^2 + v^2 - \sigma^2$. 

It is straightforward to show that $\epsilon$, $d$, and $\bold c$ defined as
follows are constants of the motion:
$$
\align
&{\aligned
   \epsilon
    :\!@!@!@!@!@!@!&= E^2(E^2 - v^2 + \sigma^2) - B^2  \\
                   &= (E^2 + \sigma^2)(E^2 - v^2) +
                                       |\bold v \times \boldsymbol \sigma|^2,
  \endaligned}
\tag13  \\
&            d :@,= e^t B,
\tag14  \\
&      \bold c :@,= e^t(E^2 \bold v + B \boldsymbol \sigma).
\tag15
\endalign
$$
(Choosing arc length for the path parameter when $\epsilon \neq 0$ restricts
the values of $\epsilon$ to 1, 0, and $-1$.)  If $d = 0$, then the particle's
scaled velocity vector $\bold v$ and spin vector $\boldsymbol \sigma$ are at
all times orthogonal to one another.  Each, if not $\bold 0$, maintains a fixed
direction in space, $\bold v$'s direction being that of the vector $\bold c$.
The particle's track through space is, therefore, a straight line in its own,
unwavering equatorial plane.  If $\bold c = \bold 0$, then $\bold v = \bold 0$,
so the particle sits in one place spinning (or not, if
$\boldsymbol \sigma = \bold 0$) and shrinking as time moves on (indeed, moving
time onward {\it by\/} shrinking).  If $d \neq 0$, then $\bold c$ is a nonzero
vector around which both $\bold v$ and $\boldsymbol \sigma$ precess, with, as
in Fig\. 3, $\bold v$ and either $\boldsymbol \sigma$ or $-\boldsymbol \sigma$
keeping $\bold c$ between them at all times (except, of course, when $\bold v$,
$\boldsymbol \sigma$ or $-\boldsymbol \sigma$, and $\bold c$ are all parallel).
 We shall see that in fact the particle in question moves on a helical track
whose axis is aligned with $\bold c$.

\epsfxsize=5truein
\epsffile[70 430 511 721]{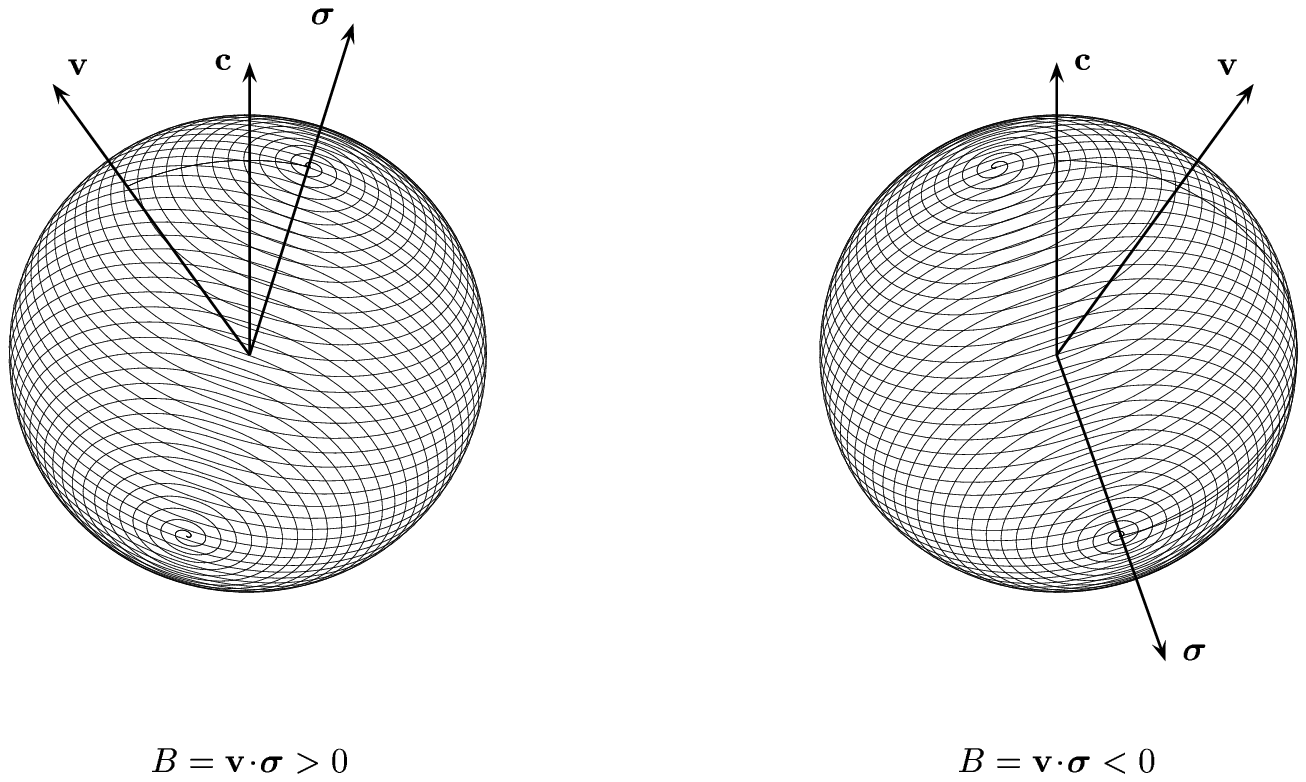}

\midspace{-35pt} \caption{Fig\. 3.   Geodesically spinning spheres with their
scaled velocity vectors $\bold v$ and their spin vectors $\boldsymbol \sigma$
precessing around the fixed vector $\bold c$.  Because $e^t B$ is a constant of
the motion, the cases $B > 0$ and $B < 0$ do not mix on a single geodesic.} 

Resolving $\bold s$, $\dot \bold s$, and $\bold v$ into their components
${\bold s}_\parallel$, ${\dot \bold s}_\parallel$, and ${\bold v}_\parallel$
parallel to $\bold c$, and ${\bold s}_\perp$, ${\dot \bold s}_\perp$, and
${\bold v}_\perp$ perpendicular to $\bold c$ allows us to express the curvature
$\kappa_\perp$ of the projection of the particle's track onto the plane through
the origin perpendicular to $\bold c$ as follows:
$$
\displaystyle {\kappa_\perp
             = {|{\dot \bold s}_\perp \times ({\dot \bold s}_\perp)\spdot|
                \over |{\dot \bold s}_\perp|^3}
             = {e^t |{\bold v}_\perp \times {\dot \bold v}_\perp|
                \over |{\bold v}_\perp|^3}}.
\tag16
$$
Some calculating then shows that
$$
\displaystyle {R_\perp := {1 \over \kappa_\perp}
                        = {(E^4 + B^2)|\bold v \times \bold c| \over c^2 |B|}},
\tag17
$$
and further that $(R_\perp)\,\dot{} = 0$.  Thus the projection, having constant
radius of curvature $R_\perp$, is a circle, and the track lies, therefore, on a
right circular cylinder whose axis is parallel to $\bold c$.  The center of
that circle, through which the axis of the cylinder must pass, is located by
the vector
$$
\displaystyle {\bold C := {\bold s}_\perp +
                             R_\perp {\bold c \times {\dot \bold s}_\perp \over
                                      |\bold c \times {\dot \bold s}_\perp|}\,
                             {\text{\tenrm sgn}} (B)},
\tag18
$$
another constant of the motion.

When $E^2 - v^2 > 0,\;= 0,\;< 0$ the particle is conventionally said to be
traveling ``slower than light, at the same speed as light, faster than light.''
In de Sitter's as in every ordinary space-time no free particle can be in one
of these states now and another later.  Here that is not the case:  a single
geodesic with $\epsilon = 1$, for example, can have $E^2 - v^2 > 0$ now,
$ = 0$ later, and $< 0$ even later.  The second of Eqs\. (13) clearly implies,
however, that if at any time the particle is traveling ``slower than light,''
then $\epsilon$ must be positive.  For this reason the geodesics on which
$\epsilon > 0$ will be taken to represent particles of nonzero inertial rest
mass.  A picture of such a particle's helical track in $\Bbb E^3$, produced
by numerical integration of the geodesic equations, is shown in Fig\. 4.

\vskip 9pt

\epsfxsize=3.5truein
\epsffile[45 406 386 721]{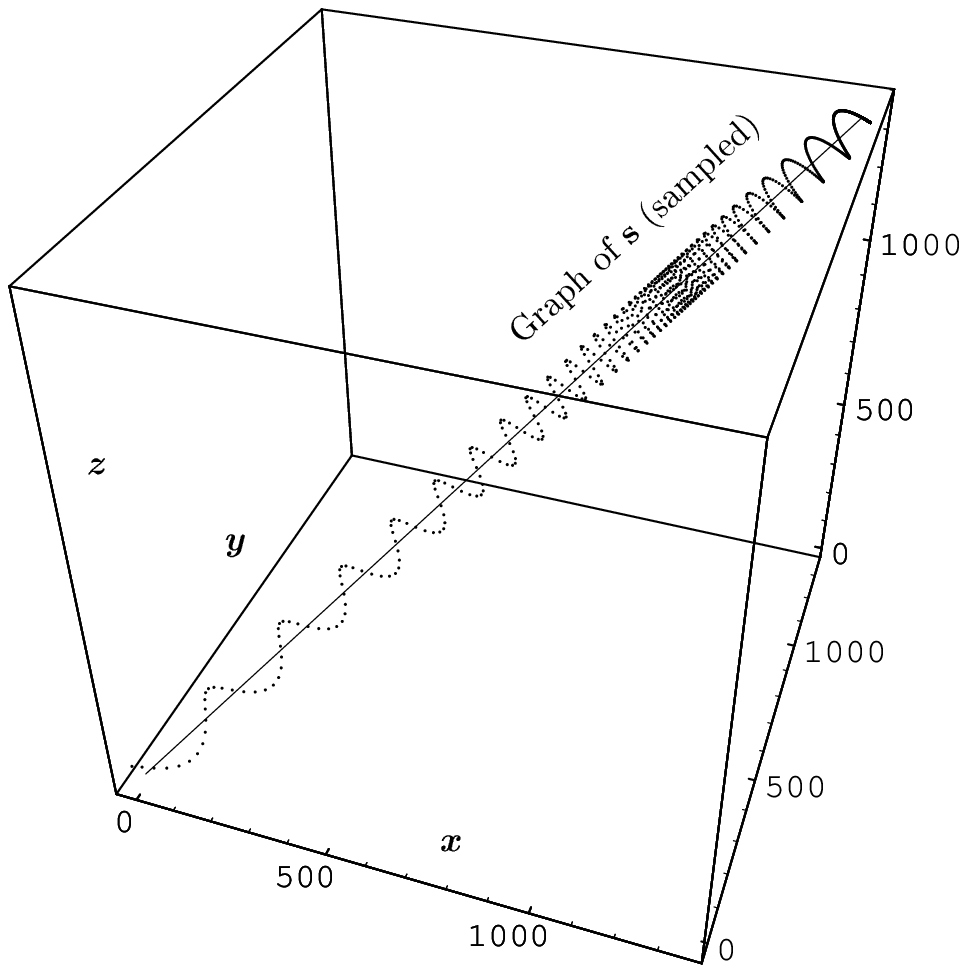}
\midspace{-0.25in} \caption{Fig\. 4.   A portion of a typical helical track of
a free spinning particle of nonzero inertial rest mass.  The parameter interval
is $[0, 70]$, sampled at intervals of .05.  The particle moves from lower left
to upper right on a cylinder of radius $R_\perp = 50$, whose axis vector
$\bold c \approx .00045 \cdot \[[\, 1, 1, 1 \,\]]$.  Some initial conditions
are $t \approx -8.78$, $E \approx .13$,
$\bold v \approx \[[ .086, -.009, .038 \,\]]$,
$\boldsymbol \sigma \approx \[[\, 5.000, 5.003, 5.001 \,\]]$,
$v_\parallel / E = .5$, $v_\perp / E = .5$, and
$v / E = \sqrt {.5}  \approx .707$ ($\text {speed of light} = 1$).}

The visible compression of the coils of the helix reflects the well-known
phenomenon that in de Sitter's universe all freely moving test particles come
asymptotically to rest at a point in space (though continuing to spread apart
as space itself expands).$^{(4)}$ The mere existence of these coils, owed
specifically to the inclusion of spin by way of the finslerian geometry, brings
to mind the quantum mechanical phenomenon of ``\text{Zitterbewegung}'' of a
spinning electron.  This ``jitter motion,'' whose existence Schr\"odinger
deduced from Dirac's relativistic wave equation,$^{(5,6)}$ is a ``microscopic''
oscillatory perturbation of the ``macroscopic'' propagation motion of the
electron.  The microscopic ``zitterspeed'' equals the speed of light, but the
``macrospeed'' is less.  In one of the manifestations of \text{Zitterbewegung}
the electron appears to follow a helical path that winds around a line
representing its macroscopic path of propagation through space.$^{(7)}$ In the
present development the quantities $v_\parallel / E$ (macrospeed) and
$v_\perp / E$ (microspeed), scaled so that $\text {lightspeed} = 1$, play roles
somewhat analogous to the macroscopic speed and the microscopic (zitter)speed
of the helical \text{Zitterbewegung} manifestation.  The de Sitter phenomenon
affects both the macrospeed, causing the compression of the coils, and the
microspeed, causing the circulatory motion (but not the spinning) to stop.
Figure 5 displays these effects explicitly, along with the variations of the
angles that the velocity $\bold v$ and the spin vector $\boldsymbol \sigma$
make with the axis vector $\bold c$ of the helix, and of the particle's
spinrate $(2 \pi)^{-1} (\sigma / E)$ (in revolutions per unit of time~$t$). 

\vskip .125in

\epsfxsize=5.875truein
\epsffile[110 450 557 671]{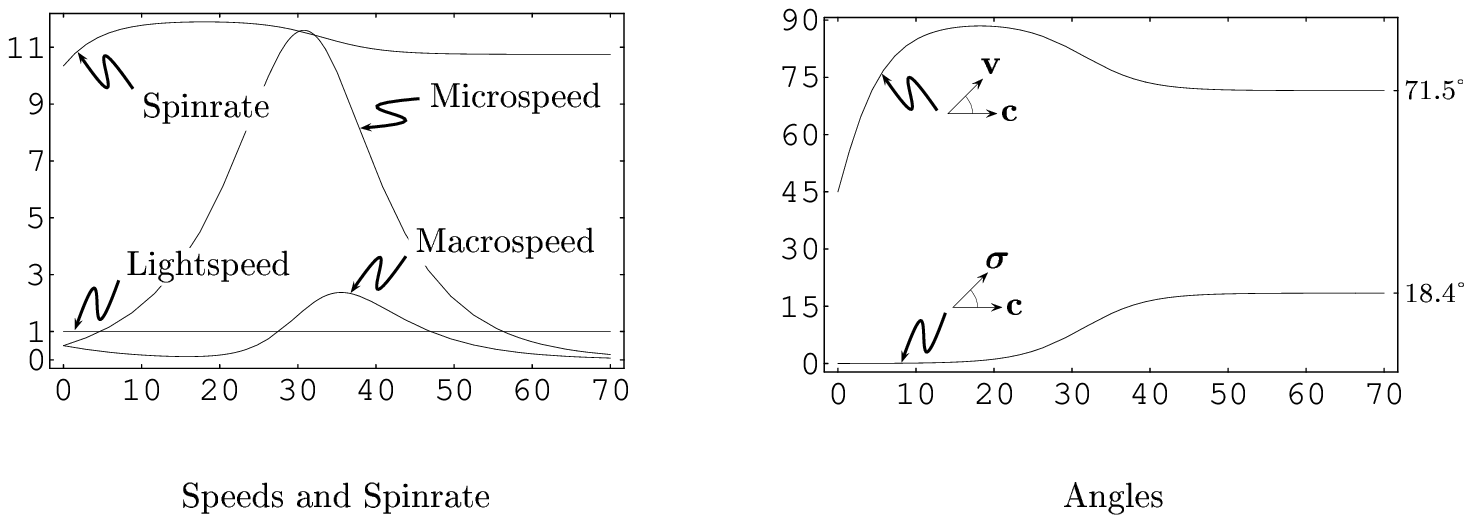}
\midspace{-.875in} \caption{Fig\. 5.   Graphs of speeds, spinrate, and angles
for the spinning particle following the helical track of Fig\. 4.  After
falling to a local minimum just above .1, the macrospeed rises to a maximum
just under 2.4 as the microspeed traverses its peak near 11.6
($\text {lightspeed} = 1$).  Initially the angles that $\bold v$ and
$\boldsymbol \sigma$ make with $\bold c$ are $45\,\degree$ and approximately
$.01\,\degree$, respectively; at the end, approximately $71.5\,\degree$ and
$18.4\,\degree$.  Their sum, the angle between $\bold v$ and
$\boldsymbol \sigma$, tends asymptotically~to~$90\,\degree$.}

The falling of the macrospeed to a local minimum produces the compression of
the helical coils seen in Fig\. 4.  Its subsequent rise to a maximum while the
microspeed is traversing its peak and the spinrate is decreasing is responsible
for the expansion of the coils after the compression.  This interplay among
kinematical variables can be interpreted as a subtle transfer of spin inertia
and orbital (micro)inertia to linear (macro)inertia as the angle from $\bold c$
to $\boldsymbol \sigma$ increases and the angle from $\bold c$ to $\bold v$
decreases.  For a clear understanding of these unusual behaviors it is
essential to remember that we are not examining motion of a pointlike particle.
Instead, we are looking at eccentric motion of a center of a spinning sphere
whose radius, according to the relation $R = e^{-t}$, is about 6495 initially,
when $t \approx -8.78$, and about 1.47 at the end, when $t \approx -.38$.
Moreover, the ``four-point'' derivation of the finslerian metric function of
Eqs\. (8) makes evident that the differential interaction of this spinning
sphere with itself, captured in the ``stationarizing'' of the finslerian arc
length integral, is an interaction taking place on the sphere itself, far from
its helically~moving~center.

If a helical track and precessing spin and velocity are typical for a free
spinning particle of nonzero rest mass, what is typical for a free spinning
particle of zero rest mass, defined as one traveling ``at the same speed as
light,'' thus on a geodesic on which $E^2 - v^2 = 0$ at all times?  For such a
particle $\epsilon$ must be 0 (one can show), and then the second of Eqs\. (13)
implies that $\bold v \times \boldsymbol \sigma = \bold 0$, hence that
$\boldsymbol \sigma$ must be parallel or antiparallel to $\bold v$.  Equations
(15) and (17) then entail that $R_\perp = 0$, thus that the track is straight.
This behavior replicates some of the behavior of spinning photons predicted by
the quantum theory of light, and does so without the aid of a Hilbert space, an
operator, a bra, or a ket.

It is both remarkable and highly suggestive that, departing from the very
elementary construction on Euclidean spheres presented here, we can a)~arrive
at a purely geometrical theory of the kinematics of free, spinning particles in
an expanding universe, b)~upon arrival, look about and find that we have
somewhat unintentionally modeled certain exotic behaviors of such particles,
behaviors first encountered in the quantum mechanical study of spinning
electrons and photons, and c)~looking back, come to suspect that we have peered
a little deeper into the mystery of time.  This short trip is perhaps in itself
a good day's journey, but it only foreshadows the labor, the pleasure, and the
satisfaction of many (maybe even {\it infinitely} many) days beyond to be spent
sailing the high seas of the Ocean of Finslerian Geometry.  For just as the
construction of the Riemannian angle, or ``two-point,'' metric of Eq\. (1) can
be extended from the manifold of spheres in Euclid's space to the manifold of
hyperspheres in Minkowski's space-time to produce a theory of
``space-time--time'' (as I outlined in Ref\. 1 and have elaborated in Ref\. 8),
the construction of the finslerian ``four-point'' metric function of Eqs\. (8)
can in direct analogy be extended from the manifold of rotated Euclidean
spheres to the manifold of Lorentz rotated Minkowskian hyperspheres to make a
theory of spinning particles in space-time-time, then further to the manifold
of rotated hyperspheres of space-time-time, and extended yet again --- time
after time after time . . .
\vskip 10pt

\noindent IN MEMORIAM.  Throughout the writing of this paper came often to mind
fond memories of Asim Orhan Barut (1926--1994), a kind and gentle spirit ever
seeking the light.

\vskip 15pt

\hfill First version:  \hskip 18.5pt March, 1995
\vskip -3pt
\hfill Revised:  September, 1995
\vskip -3pt
\hfill Revised:  \hskip 11pt October, 1995
\vskip -3pt
\hfill Revised:  \hskip 11pt October, 1996

\vskip 10pt

\noindent Homer G. Ellis\newline
Department of Mathematics\newline
University of Colorado at Boulder\newline
395 UCB\newline
Boulder, Colorado  80309-0395\newline\newline
Telephone: (303) 492-7754 (office); (303) 499-4027 (home)\newline
Email: ellis\@euclid.colorado.edu\newline
Fax: (303) 492-7707

\enddocument